\documentclass[10pt,letterpaper,twocolumn]{article}
\usepackage[margin=0.75in]{geometry}
\usepackage{times}
\usepackage[round,authoryear]{natbib}
\usepackage{hyperref}
\usepackage{graphicx} 
\usepackage{amsmath} 
\usepackage[table]{xcolor}
\usepackage{makecell}
\usepackage{booktabs}
\usepackage{longtable}
\usepackage{array}
\usepackage[most]{tcolorbox}
\usepackage{fvextra}
\usepackage{natbib} 
\usepackage{caption} 
\usepackage{algorithm}
\usepackage{multirow}
\usepackage{algorithmic}

\usepackage{newfloat}
\usepackage{listings}
\DeclareCaptionStyle{ruled}{labelfont=normalfont,labelsep=colon,strut=off} 
\floatstyle{ruled}
\newfloat{listing}{tb}{lst}{}
\floatname{listing}{Listing}

\newtcolorbox{promptbox}{
    enhanced,
    breakable,
    colback=gray!8,
    colframe=gray!35,
    boxrule=0.4pt,
    arc=3mm,
    left=4mm,
    right=4mm,
    top=3mm,
    bottom=3mm,
    drop shadow,
    before skip=6pt,
    after skip=14pt
}

\title{
ASCon: A Direction-Aware Reciprocal Agent--Step Contextualization Model
for Failure Attribution in Multi-Agent Systems
}

\author{
Shuyu Jiang$^{1,2}$,
Yue Ran$^{1}$,
Kaiyu Xu$^{1}$,
 Xingshu Chen $^{1}$,
 Yi Zhang$^{1}$,
 Hao Ren $^{1}$,
 Rui Tang$^{1}$,
 Tianwei Zhang$^{2}$
\\[6pt]
 $^{1}$Sichuan University, China
 \quad\quad
 $^{2}$Nanyang Technological University, Singapore
}
\date{} 

\begin{document}

\maketitle
\begin{abstract}
Failure attribution in LLM-based multi-agent systems (MAS) aims to answer who caused failures, when they occurred, and why by identifying responsible targets including faulty agents, erroneous steps, and failure modes. Existing methods have primarily focused on developing dedicated models for specific attribution targets, with limited attention to the evidential dependencies among them. Despite these attribution targets are different, they rely on common diagnostic evidence from MAS trajectories, including task constraints, agent roles, behavioral histories and inter-agent interactions. This commonality motivates us to develop a unified representation model that aggregates the trajectory evidence into individual agent and step representations, which can subsequently be adapted to different attribution targets. Accordingly, we propose ASCon, a direction-aware reciprocal \textbf{A}gent--\textbf{S}tep \textbf{Con}textualization model for multiple failure attribution targets. ASCon introduces direction-aware graph attention to model execution context, masked step-to-agent attention to construct behavior-aware agent representations, and agent-conditioned step contextualization to incorporate agent context back into step representations. The resulting contextualized representations enable different attribution targets through lightweight target-specific heads. Experiments show that ASCon can improve faulty-agent detection by 5.83\%+ in micro-accuracy, faulty-step detection by 10.63\%+ in micro-accuracy, and failure-mode detection by 14.73\%+ in Macro-F1. Meanwhile, it can also substantially enhance the LLM-based methods' attribution capabilities in out-of-domain scenarios. Code is available at: \href{https://github.com/Shuyu-07/ASCon}{https://github.com/Shuyu-07/ASCon}
\end{abstract}
\section{Introduction}
Large language model (LLM)-based multi-agent systems (MAS) have emerged as a promising paradigm for complex problem solving by coordinating agents with specialized roles and tool-use capabilities across tasks such as software development, information retrieval, and scientific analysis~\citep{fourney2024magenticone,ghareeb2026multi,qian2024chatdev}. Meanwhile, their growing capabilities are also accompanied by more complex failure modes like inter-agent misalignment, making system debugging and error correction increasingly difficult~\citep{qi2026beyond,yu2025netsafe}.
Diagnosing such failures requires developers to inspect lengthy traces with domain knowledge, often taking several to tens of minutes per instance~\citep{cemri2025why, epperson2025interactive, zhang2025who&when}.
This substantial diagnostic burden made automated failure attribution increasingly important for scalable MAS debugging, maintenance, and subsequent optimizations~\citep{ma2026dover,wang2026xagen,in2026rethinking}.

\begin{figure}[t!]
    \centering
    \includegraphics[width=1.\linewidth]{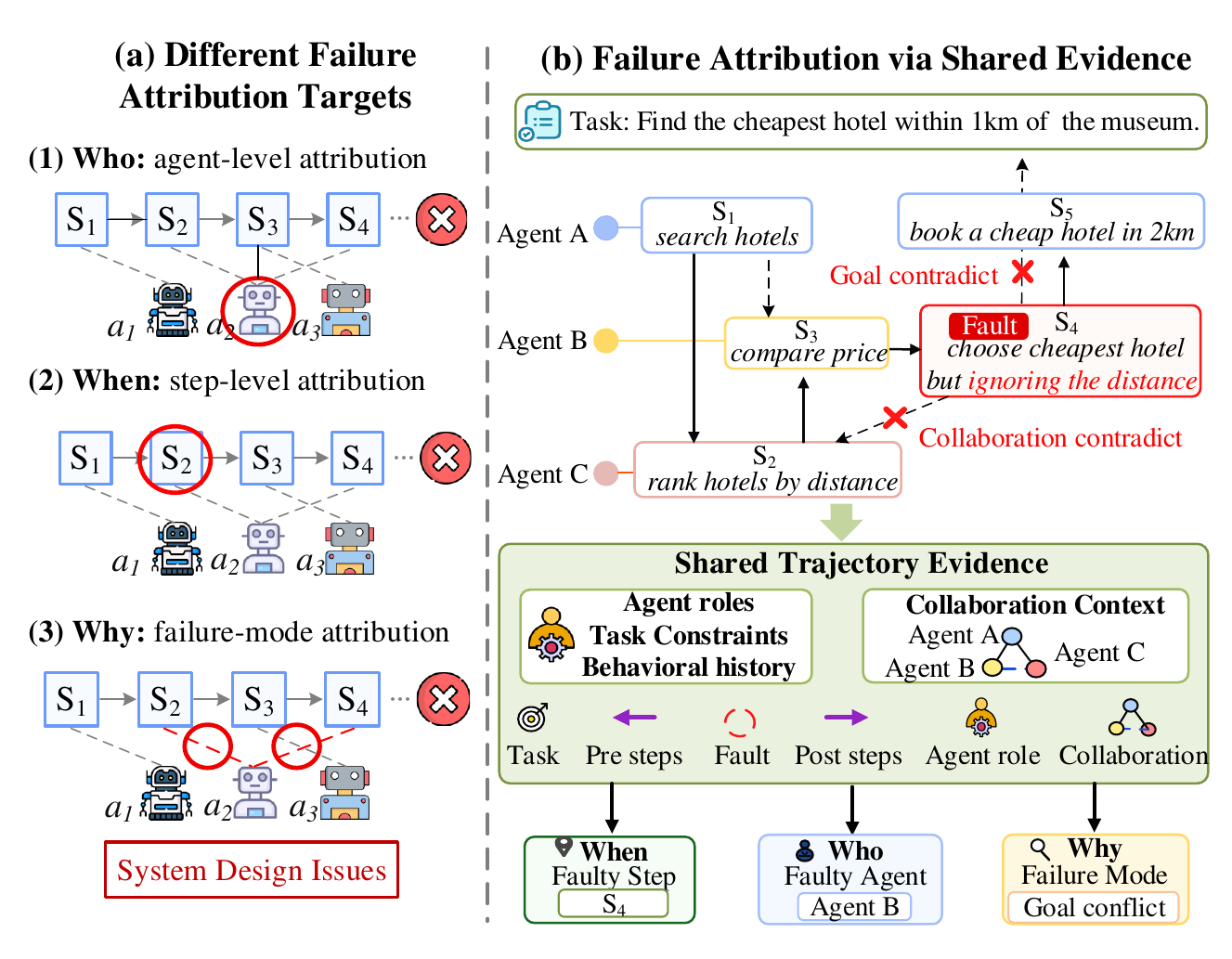}
    \caption{(a) MAS failure attribution are commonly formulated at separate agent, step, and failure-mode levels. (b) These targets share the same trajectory evidence, which ASCon models for unified attribution across all three levels. }
    \label{fig:intro}  
\end{figure}

As shown in Figure~\ref{fig:intro}, MAS failure attribution refers to the identification of key failure factors from an execution trace, including: \emph{who} -- the agents responsible for the failure; \emph{when} -- the steps at which the failure occurs or first manifests; and \emph{why} -- the corresponding failure modes or error modes~\citep{liu2026pro}. Existing studies typically develop separate methods for different attribution targets. For step-level attribution, failures are usually located through LLM-based search and ranking, or by modeling anomalous temporal changes in behavioral sequences~\citep{zhang2025who&when, zhu2026raffles, sun2026scope}. Agent-level methods instead determine responsibility by examining whether an agent's behavioral history is consistent with its intended goal or assigned role~\citep{geng2026failure,wu2026intention}. Failure-mode attribution methods commonly abstract these anomalous behaviors into predefined error categories through LLM inference~\citep{kong2026aegis}.

Although agent, step, and failure-mode attribution are often formulated separately in these valuable studies, they draw on closely related evidence from the same execution trajectory.
For example, whether a candidate step is faulty is determined from itself as well as its preceding and subsequent execution actions, the task constraints, and the role of the executing agent~\citep{qiao2026verifymas,chen2026traceelephant}. Likewise, attributing responsibility to an agent requires aggregating evidence from its historical actions and examining whether they remain consistent with its role and the task constraints~\citep{ge2026spectrum}. On this basis, failure-mode identification further characterizes how the observed anomalous behavior deviates from the exception~\citep{kong2026aegis}. Therefore, task constraints, agent role settings,  behavioral history, and inter-agent interactions jointly provide the contextual evidence for different failure attribution targets.

Motivated by this observation, we propose \textbf{ASCon}, a direction-aware reciprocal agent-step contextualization model for multi-agent failure attribution, which unifies step-, agent-, and failure-mode attribution within a single framework. ASCon first introduces a direction-aware graph attention to separately model the preceding and subsequent execution contexts. It then uses masked step-to-agent attention to aggregate the steps performed by each agent into a behavior-aware agent representation, which is further contextualized over an agent interaction graph to capture cross-agent collaboration. Finally, the agent representations are injected back into their corresponding step representations, forming a reciprocal contextualization process between steps and agents. Through this reciprocal information flow, ASCon integrates task constraints, behavioral context, and collaboration patterns into agent and step representations, thereby enabling flexible adaptation to different supervision schemes through lightweight prediction heads.

We evaluate ASCon on TracerTraj and Aegis-Bench, which together cover agent-level, step-level, and failure-mode attribution. Experimental results show that ASCon achieves improvements of 5.83\%+ in agent-level micro-accuracy, 10.63\%+ in step-level micro-accuracy, and 14.73\%+  in failure-mode macro-F1. Moreover, when incorporated into LLM-based attribution methods, ASCon improves out-of-domain agent-level micro-accuracy by 8.69\%+, step-level micro-accuracy by 12.50\%+ and failure-mode-level macro-F1 by 3.15\%+.

Our main contributions are as follows:
\begin{itemize}
    \item We propose ASCon, a reciprocal agent-step contextualization framework for multi-agent failure attribution, which supports agent-, step-, and failure-mode attribution under heterogeneous supervision settings.
    \item  We introduce direction-aware graph attention for modeling preceding and subsequent step dependencies, and masked step-to-agent attention for selectively aggregating agent-specific behavioral evidence.
    \item Comprehensive experiments indicate that ASCon consistently improves performance across attribution targets and effectively enhances out-of-domain generalization.
\end{itemize}

\section{Related Works}
MAS failure attribution aims to identify responsible agents, erroneous steps, or failure modes from execution trajectories~\citep{zhang2025who&when, chen2026traceelephant,kong2026aegis}. Existing methods can be broadly divided into prompt-based and learning-based failure attributions.

\paragraph{Prompt-based methods.} Prompt-based  methods directly prompt LLMs to inspect execution trajectories and localize suspicious behaviors without training a dedicated attribution model. Specifically, Who\&When studies All-at-Once, step-by-step, and binary-search strategies~\citep{zhang2025who&when}; SDBL narrows the candidate scope before localization~\citep{sun2026scope}; and RAFFLES iteratively refines fault hypotheses through judge-evaluator reasoning~\citep{zhu2026raffles}. Other studies incorporate counterfactual analysis or causal structures to guide the reasoning process~\citep{west2025a2p,wang2026chief,ge2026spectrum}. While these methods offer flexible trajectory analysis, their attribution quality is largely determined by the reasoning capability of the underlying LLM, especially for long and structurally complex traces.

\paragraph{Learning-based methods.} Learning-based methods construct task-specific supervision and learn representations of failed trajectories. For instance, AgentTracer generates annotated traces through counterfactual replay and programmed fault injection, and trains a dedicated failure tracer~\citep{zhang2026agentracer}. Aegis introduces context-aware error generation and annotations of faulty agents and error modes, supporting supervised fine-tuning and reinforcement learning~\citep{kong2026aegis}.  StepFinder models trajectories as temporal semantic sequences for step-level localization~\citep{zhu2026stepfinder}, and reduces reliance on inference-time reasoning by learning attribution patterns directly from execution data.

\begin{figure*}[!t]
    \centering
    \includegraphics[width=.95\linewidth]{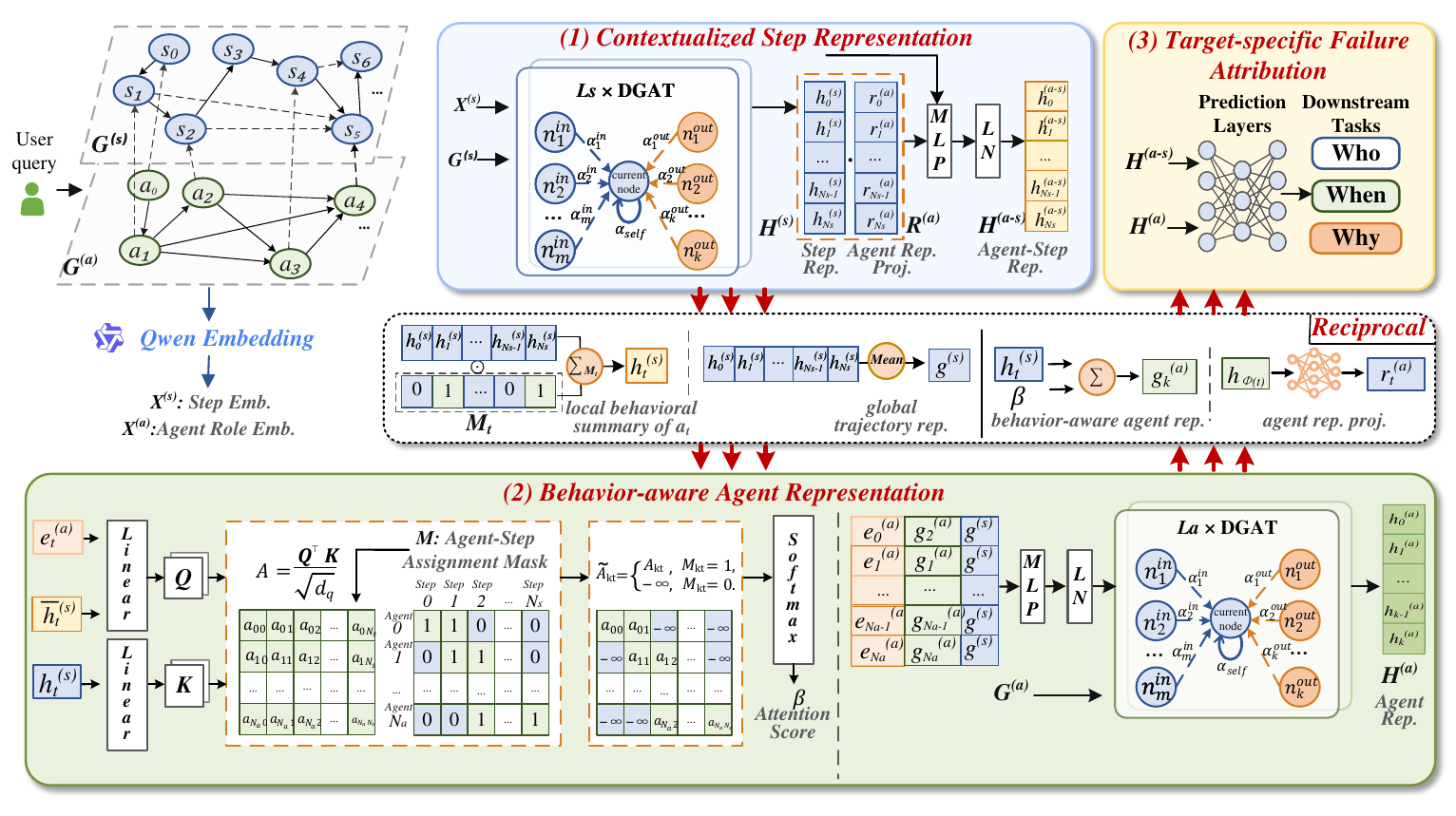}
    \caption{The framework of our ASCon model. 
    }
    \label{fig2}
\end{figure*}

\section{Problem Formulation}
\label{sec:problem}
An LLM-based multi-agent system consists of a set of agents
$\mathcal{A}=\{a_1,a_2,\ldots,a_{N_a}\}$ that collaborate to complete a task.
Its execution is recorded as a trajectory
$\tau=\{s_1,s_2,\ldots,s_{N_s}\}$, where each step is represented as $s_t=(t,\phi(t),c_t)$.
Here, $t$ is the step index, $\phi(t)\in\{1,\ldots,N_a\}$ identifies the agent executing the step, and $c_t$ denotes its textual execution record, such as an agent message, intermediate reasoning, tool invocation, or tool feedback.

Given a failed trajectory $\tau$, MAS failure attribution aims to determine
\emph{who} is responsible for the failure, \emph{when} the failure occurs, and \emph{why} it occurs.
Let $\mathcal{Y}=\{m_1,m_2,\ldots,m_M\}$ denote the set of failure modes.
The ground-truth attribution is represented as
$\mathcal{G}(\tau)
=\left\{ \left(a_i^{*},\mathcal{S}_i^{*},\mathcal{Y}_i^{*}\right)
\right\}_{i=1}^{K}$,
where $a_i^{*}\in\mathcal{A}$ is a faulty agent,
$\mathcal{S}_i^{*}\subseteq\tau$ contains its faulty steps, and
$\mathcal{Y}_i^{*}\subseteq\mathcal{Y}$ contains the corresponding failure modes.
The objective is to learn an attribution function $f_{\theta}$ that maps the failed trajectory to a structured prediction: $f_{\theta}:\tau\mapsto\widehat{\mathcal{G}}(\tau)
\approx \mathcal{G}(\tau)$. Accordingly, the predicted agents, steps, and failure modes constitute the
\emph{who}, \emph{when}, and \emph{why} of MAS failures.

\section{Methodology}
As illustrated in Figure~\ref{fig2}, ASCon learns mutually contextualized representations of steps and agents from an MAS execution trajectory. 
It first introduces a direction-aware graph attention mechanism that separately captures preceding and succeeding context in both step execution and agent interaction structures. 
Then it contextualizes individual steps along the execution trajectory, aggregates their behavioral evidence to construct agent representations, and conditions each step on the representation of its executing agent. 
Finally, target-specific prediction layers map the resulting agent and agent-conditioned step representations to the required failure-attribution outputs.

\subsection{Direction-aware Graph Attention}
To capture the evolving dependencies in MAS execution, where earlier actions influence subsequent decisions and agent behaviors are coordinated through interactions, ASCon models each trajectory as a directed step graph and a directed agent graph. 
For the step graph, the user query is prepended as step $s_0$ to provide task constraints for evaluating subsequent behaviors, yielding the augmented execution sequence $\widetilde{\mathcal{S}}=(s_0,s_1,\ldots,s_{N_s}).$
Each step in $\widetilde{\mathcal{S}}$ is treated as a graph node, forming the node set
$V^{(s)}=\{s_0,s_1,\ldots,s_{N_s}\}.$ Adjacent steps in the execution sequence are connected according to their temporal order, forming the directed step graph $G^{(s)}=(V^{(s)},\mathcal{E}^{(s)})$, 
where $\mathcal{E}^{(s)}=\{(s_t,s_{t+1})\mid 0\leq t<N_s\}$. Each directed edge represents a direct execution dependency between consecutive actions, preserving the temporal flow of the trajectory. For the agent graph, ASCon introduces a user node $a_0$ corresponding to $s_0$ with $\phi(0)=0$. The augmented agent node set is defined as $V^{(a)}=\widetilde{\mathcal{A}}=\{a_0,a_1,\ldots,a_{N_a}\}$. Agents responsible for adjacent steps are connected following the same execution direction, yielding the directed agent graph $G^{(a)}=(V^{(a)},\mathcal{E}^{(a)})$, where $\mathcal{E}^{(a)}
=
\{(a_{\phi(t)},a_{\phi(t+1)})\mid 0\leq t<N_s\}$.

In failure attribution, the preceding and succeeding node neighbors in $G^{(s)}$ and $G^{(a)}$ provide asymmetric evidence: the former provide the context for assessing whether a node is anomalous, whereas the latter reveal its potential downstream effects. Treating both directions identically may therefore blur the distinction between a root failure and its consequences. Thus, ASCon introduces a direction-aware graph attention network (DGAT) that separately models incoming, outgoing, and self information.

Let $G=(\mathcal{V},\mathcal{E})$ uniformly denote the constructed \(G^{(s)}\) or \(G^{(a)}\), and $\mathbf{u}_i$ denote the input feature of node $v_i$. 
Its incoming and outgoing neighborhoods are defined as
$\mathcal{N}_{\mathrm{in}}(i)
=
\{v_j\mid(v_j,v_i)\in\mathcal{E}\}$, and
$\mathcal{N}_{\mathrm{out}}(i)
=
\{v_j\mid(v_i,v_j)\in\mathcal{E}\}$.
At the \(l\)-th layer, DGAT first computes attention scores for incoming and outgoing edges separately as follows:
\begin{equation}
\begin{aligned}
e_{ji,\mathrm{in}}^{l}
&=
\mathrm{LeakyReLU}
\left(
{\mathbf{a}_{\mathrm{in}}^{l}}^{\top}
\left[
\mathbf{W}_{\mathrm{in}}^{l}\mathbf{u}_i^{l};
\mathbf{W}_{\mathrm{in}}^{l}\mathbf{u}_j^{l}
\right]
\right),\\
e_{ij,\mathrm{out}}^{l}
&=
\mathrm{LeakyReLU}
\left(
{\mathbf{a}_{\mathrm{out}}^{l}}^{\top}
\left[
\mathbf{W}_{\mathrm{out}}^{l}\mathbf{u}_i^{l};
\mathbf{W}_{\mathrm{out}}^{l}\mathbf{u}_j^{l}
\right]
\right).
\end{aligned}
\end{equation}
\(\mathbf{W}_{\mathrm{in}}^{l}\) and \(\mathbf{W}_{\mathrm{out}}^{l}\) are  learnable direction-specific transformation matrices. \(\mathbf{a}_{\mathrm{in}}^{l}\) and \(\mathbf{a}_{\mathrm{out}}^{l}\) are learnable attention vectors.

The attention scores are normalized separately along the incoming and outgoing directions, and the direction-specific messages $\mathbf{m}_{i}$ are aggregated by taking an attention-weighted sum of the corresponding neighborhoods as follows:
\begin{equation}
\begin{aligned}
\mathbf{m}_{i,\mathrm{in}}^{l}
&=
\sum_{v_j\in\mathcal{N}_{\mathrm{in}}(i)}
\alpha_{ji,\mathrm{in}}^{l}
\mathbf{W}_{\mathrm{in}}^{l}\mathbf{u}_j^{l},\\
\mathbf{m}_{i,\mathrm{out}}^{l}
&=
\sum_{v_j\in\mathcal{N}_{\mathrm{out}}(i)}
\alpha_{ij,\mathrm{out}}^{l}
\mathbf{W}_{\mathrm{out}}^{l}\mathbf{u}_j^{l}.
\end{aligned}
\end{equation}

\begin{equation}
\begin{aligned}
\alpha_{ji,\mathrm{in}}^{l}
&=
\operatorname{softmax}_{j\in\mathcal{N}_{\mathrm{in}}(i)}
(e_{ji,\mathrm{in}}^{l}),\\
\quad
\alpha_{ij,\mathrm{out}}^{l}
&=
\operatorname{softmax}_{j\in\mathcal{N}_{\mathrm{out}}(i)}
(e_{ij,\mathrm{out}}^{l}).
\end{aligned}
\end{equation}

For the next-layer, the node representation is updated by combining self, incoming, and outgoing information:
\begin{equation}
\widetilde{\mathbf{u}}_i^{l+1}=\text{ELU}\left(
\mathbf{W}_{\mathrm{self}}^{l}\mathbf{u}_i^{l}
+\mathbf{m}_{i,\mathrm{in}}^{l}
+\mathbf{m}_{i,\mathrm{out}}^{l}
\right),
\end{equation}
where \(\mathbf{W}_{\mathrm{self}}^{l}\) is the self-transformation matrix, and \(\text{ELU}(\cdot)\) is the nonlinear activation function.

\subsection{Contextualized Step Representation}
Since the correctness of steps depends not only on its own content but also on its consistency with task constraints and surrounding actions, ASCon learns contextualized step representations over the directed step graph $G^{(s)}$, whose edges record inter-step execution dependencies. Each node in $\widetilde{\mathcal{S}}$ is first encoded into a semantic embedding $\mathbf{e}_t^{(s)}= f_{\mathrm{LM}}(c_t)$,
where $f_{\mathrm{LM}}$ denotes a pretrained text encoder and $c_t$ is the step content. Then, ASCon applies $L_s$ DGAT layers to \(G^{(s)}\) and its step embeddings $\mathbf{X}^{(s)}
=[\mathbf{e}_0^{(s)};
\mathbf{e}_1^{(s)};
\ldots;
\mathbf{e}_{N_s}^{(s)}]$  to get the contextualized step representation $\mathbf{H}^{(s)}$ as follows: 
\begin{equation}
\mathbf{H}^{(s)}
=
\mathrm{DGAT}_{L_s}^{(s)}
\left(
\mathbf{X}^{(s)},G^{(s)}
\right),
\end{equation}
where $\mathbf{h}_t^{(s)}$ is the contextualized representation of step $s_t$.

\paragraph{Agent-conditioned step representation.}
Beyond the task constraints and surrounding actions encoded above, the correctness of a step also depends on its consistency with the setting of the executing agent. Therefore, ASCon constructs an agent-conditioned step representation by incorporating the behavioral state of the responsible agent.
For step \(s_t\), let \(a_{\phi(t)}\) be the agent associated with this step. 
The behavior-aware agent representation \(\mathbf{h}_{\phi(t)}^{(a)}\), computed in the next subsection, is projected into the step representation space: 
\begin{equation}
\mathbf{r}_t^{(a)}
=
\mathbf{W}_{a2s}\mathbf{h}_{\phi(t)}^{(a)}.    
\end{equation}
The agent-conditioned step representation is computed as 
\begin{equation}
\mathbf{h}_t^{(a-s)}=\mathrm{LayerNorm}
\left(
\mathrm{MLP}
\left(
\left[
\mathbf{h}_t^{(s)};
\mathbf{h}_t^{(s)}\odot\mathbf{r}_t^{(a)}
\right]
\right)
\right),
\end{equation}
where $\odot$ denotes element-wise multiplication.

\begin{table*}[thbp!]
\centering
\fontsize{10pt}{10.5pt}\selectfont
\begin{tabular*}{0.95\textwidth}{@{\extracolsep{\fill}} l c c c c c c | c c c c}
\toprule
\multirow{2}{*}{\textbf{Method}} &
\multicolumn{2}{c}{\textbf{Code}} &
\multicolumn{2}{c}{\textbf{Math}} &
\multicolumn{2}{c|}{\textbf{Agentic}} &
\multicolumn{2}{c}{\textbf{Micro-Accuracy}} &
\multicolumn{2}{c}{\textbf{Macro-Accuracy}} \\
\cmidrule(lr){2-3}
\cmidrule(lr){4-5}
\cmidrule(lr){6-7}
\cmidrule(lr){8-9}
\cmidrule(lr){10-11}
& \textbf{Agent} & \textbf{Step}
& \textbf{Agent} & \textbf{Step}
& \textbf{Agent} & \textbf{Step}
& \textbf{Agent} & \textbf{Step}
& \textbf{Agent} & \textbf{Step} \\
\midrule

DeepSeek-V4-Pro
& 33.07
& 11.81
& 53.97
& 25.40
& 58.83
& 37.83
& 54.30
& 32.66
& 48.62
& 25.01 \\

Gemini-3.1-Flash
& 74.80
& 24.41
& 50.79
& 42.86
& 60.50
& 21.00
& 62.03
& 23.29
& 62.03
& 29.42 \\

All-at-Once
& 66.14
& 7.87
& 34.92
& 30.16
& 61.00
& 22.17
& 59.75
& 20.51
& 54.02
& 20.07 \\

Step-by-Step
& 65.35
& 10.24
& 57.14
& 36.51
& 69.67
& 23.50
& 67.98
& 22.41
& 64.05
& 23.42 \\

Binary Search
& 48.82
& 2.36
& 42.86
& 28.57
& 53.00
& 20.50
& 51.52
& 18.23
& 48.23
& 17.14 \\

SDBL (EASD)
& 70.08
& 7.09
& 44.44
& 38.10
& 64.83
& 35.67
& 64.05
& 31.27
& 59.78
& 26.95 \\


Qwen-SFT
& 44.09
& 5.51
& 49.21
& 22.22
& 61.33
& 38.00
& 57.59
& 31.52
& 51.54
& 21.91 \\

AgentTracer
& \underline{83.46}
& 7.09
& 36.51
& 42.86
& 61.00
& 23.50
& 62.66
& 22.41
& 60.32
& 24.48 \\

StepFinder
& 81.89
& \underline{41.73}
& \underline{\textbf{79.37}}
& \underline{\textbf{74.60}}
& \underline{73.33}
& \underline{52.67}
& \underline{75.18}
& \underline{52.66}
& \underline{78.20}
& \underline{56.33} \\
\midrule

\textbf{ASCon}
& \textbf{85.04}
& \textbf{52.76}
& \textbf{79.37}
& \textbf{74.60}
& \textbf{80.33}
& \textbf{64.33}
& \textbf{81.01}
& \textbf{63.29}
& \textbf{81.58}
& \textbf{63.90} \\
\bottomrule
\end{tabular*}
\caption{Agent- and step-level attribution accuracy (\%) on the root-fault attribution task. Best results are in \textbf{bold}, with best baselines \underline{underlined}. Columns 2$\sim$7 report detailed sub-test set results and the last four columns report overall average accuracies.}
\label{tab:root-fault}
\end{table*}

\subsection{Behavior-aware Agent Representation}
After step-level encoding, ASCon lifts behavioral evidence from steps to agents. Since an agent is characterized by its executed steps and different steps contribute unequally to fault diagnosis, we introduce masked step-to-agent attention to restrict aggregation to the agent's own steps and emphasize the most informative ones.

\paragraph{Masked step-to-agent attention.} To incorporate the task constraint into the agent aggregation, we introduce a user agent $a_0$ associated only with step $s_0$. 
The augmented agent node set is
$\widetilde{\mathcal{A}}=\{a_0,a_1,\ldots,a_{N_a}\}$.  Let \(\mathbf{e}_k^{(a)}\) denote the role embedding of agent \(a_k\), where \(0\leq k\leq N_a\). 
We define an agent-step assignment matrix 
\(\mathbf{M}\in\{0,1\}^{(N_a+1)\times (N_s+1)}\), where \(M_{kt}=1\) if step \(s_t\) is associated with agent \(a_k\), and \(M_{kt}=0\) otherwise. 
For each agent \(a_k\), ASCon first computes a local behavioral summary from its associated steps:
\begin{equation}
\bar{\mathbf{h}}_k^{(s)}
=
\frac{
\sum_{t=0}^{N_s} M_{kt}\mathbf{h}_t^{(s)}
}{
\sum_{t=0}^{N_s} M_{kt}
}.
\end{equation}
This summary provides an agent-specific context for constructing the pooling query. 
Then, the query for step-to-agent attention \(a_k\) is generated as 
\(\mathbf{q}_k=\mathbf{W}_{q}[\mathbf{e}_k^{(a)};\bar{\mathbf{h}}_k^{(s)}]\), 
while each step representation is projected into a key vector 
\(\mathbf{k}_t=\mathbf{W}_{K}\mathbf{h}_t^{(s)}\). 
The agent-step relevance is then measured by scaled dot-product attention as
\(A_{kt}=\mathbf{q}_k^{\top}\mathbf{k}_t/\sqrt{d_q}\).

The assignment mask further restricts each agent only attention to its own steps as follows:
\begin{equation}
\beta_{kt}=\operatorname{softmax}_{t}(\widetilde{A}_{kt}),  \\
\widetilde{A}_{kt}
=
\begin{cases}
A_{kt}, & M_{kt}=1,\\
-\infty, & M_{kt}=0,
\end{cases}
\end{equation}
where $\beta_{kt}$ is the attention score.
The behavior-aware representation of agent $k$ is then obtained by aggregating its step-level evidence: $\mathbf{g}_k^{(a)}
=\sum_{t=0}^{N_s} \beta_{kt}\mathbf{h}_t^{(s)}$.

\paragraph{Agent graph contextualization.} 
To complement agent-specific evidence with the overall task context, ASCon further computes a global trajectory representation $\mathbf{g}^{(s)}$ by mean-pooling all step representations.
The initial agent representations are formed by combining agent role, behavior-aware evidence, and global trajectory context:
\begin{equation}
\mathbf{z}_k^{(a)}
=
\mathrm{LayerNorm}
\left(
\mathrm{MLP}
\left(
[
\mathbf{e}_k^{(a)};
\mathbf{g}_k^{(a)};
\mathbf{g}^{(s)}
]
\right)
\right).
\end{equation}
Similarly, ASCon applies \(L_a\) DGAT layers on the agent graph \(G^{(a)}\) to learn the final agent representation $\mathbf{H}^{(a)}$, as follows:
\begin{equation}
\mathbf{H}^{(a)}=
\mathrm{DGAT}_{L_a}^{(a)}
\left(
\mathbf{Z}^{(a)},G^{(a)}
\right),
\end{equation}
where \(\mathbf{Z}^{(a)}=[\mathbf{z}_0^{(a)};\mathbf{z}_1^{(a)};\ldots;\mathbf{z}_{N_a}^{(a)}]\), and \(\mathbf{h}_k^{(a)}\in \mathbf{H}^{(a)} \) serves as a contextual anchor for the failure attribution task.
\subsection{Target-specific Failure Attribution}
The contextualized step and agent representations learned above provide a common basis for identifying faulty agents (\emph{who}), faulty steps (\emph{when}), and failure modes (\emph{why}).
Specifically, agent representations support faulty agent and failure mode identification, agent-conditioned step representations support faulty step localization.
Generally, existing MAS failure-attribution studies primarily instantiate these targets in two representative forms: (a) root-fault attribution, which identifies one root responsible agent and one key failure step, and (b) failure-mode attribution, which identifies faulty agents and their corresponding failure modes.
Accordingly, ASCon employs the following two prediction heads for these two settings.

\paragraph{Root-fault attribution.}
For root-fault attribution, ASCon independently predicts whether each agent and step is faulty using two binary classification heads. The fault probabilities are computed as
$\hat{y}_k^{(a)}=\sigma(\mathrm{MLP}_{a}(\mathbf{h}_k^{(a)}))$
and
$\hat{y}_t^{(s)}=\sigma(\mathrm{MLP}_{s}(\mathbf{h}_t^{(a-s)}))$,
where $\sigma(\cdot)$ denotes the sigmoid function.
Let $\hat{\mathbf{y}}^{(a)}$ and $\hat{\mathbf{y}}^{(s)}$ collect the predicted probabilities of all agents and steps, with $\mathbf{y}^{(a)}$ and $\mathbf{y}^{(s)}$ denoting their binary labels. The training objective is
\begin{equation}
\mathcal{L}_{\mathrm{r}}
=
\sum\nolimits_{q\in\{a,s\}}
\mathrm{BCE}
\left(
\hat{\mathbf{y}}^{(q)},
\mathbf{y}^{(q)}
\right).
\end{equation}
During inference, the highest-probability agent and step are selected as the attribution result.

\paragraph{Failure-mode attribution.}
As the relevant diagnostic evidence of an agent's failure mode is distributed across its steps, ASCon combines each contextualized agent representation with associated steps.
For agent $a_k$, the step-level evidence is pooled as
$\bar{\mathbf{h}}_k^{(a-s)}
=
\sum_{t=1}^{N_s}M_{kt}\mathbf{h}_t^{(a-s)}
/
\sum_{t=1}^{N_s}M_{kt}$
and fused with its agent representation:
\begin{equation}
\mathbf{o}_k^{(a)}
=
\mathrm{LayerNorm}
(
\mathrm{MLP}_{f}
(
[
\mathbf{h}_k^{(a)};
\bar{\mathbf{h}}_k^{(a-s)}
]
)).
\end{equation}
ASCon then predicts the faulty-agent probability
$\hat{y}_k^{(a)}
=
\sigma(\mathrm{MLP}_{a}(\mathbf{o}_k^{(a)}))$
and failure-mode distribution
$\hat{\mathbf{b}}_k^{(m)}
=
\operatorname{softmax}(\mathrm{MLP}_{m}(\mathbf{o}_k^{(a)}))$.
Let $\hat{\mathbf{y}}^{(a)}$ denote the faulty-agent probabilities and $\mathbf{y}^{(a)}$ binary labels, while $\hat{\mathbf{B}}_{\mathcal{F}}^{(m)}$ and $\mathbf{b}_{\mathcal{F}}^{(m)}$ denote the predicted distributions and labels of the faulty agents in set $\mathcal{F}=\{k\mid y_k^{(a)}=1\}$. 
The training objective is
\begin{equation}
\mathcal{L}_{\mathrm{m}}
=
\mathrm{BCE}
\left(
\hat{\mathbf{y}}^{(a)},
\mathbf{y}^{(a)}
\right)
+
\mathrm{CE}
\left(
\hat{\mathbf{B}}_{\mathcal{F}}^{(m)},
\mathbf{b}_{\mathcal{F}}^{(m)}
\right).
\end{equation}

\section{Experiments and Analysis}

\subsection{Experimental Settings}
\paragraph{Datasets.}
For root-fault and failure-mode attributions, we select \textbf{TracerTraj}~\citep{zhang2026agentracer} and \textbf{Aegis-Bench}~\citep{kong2026aegis} as the corresponding evaluation benchmarks.
TracerTraj contains trajectories generated by six MASs across coding, mathematical reasoning, and general agentic tasks. We use its extended version as the benchmark, which comprises 3,208 training and 790 test trajectories (127 code, 63 math, and 600 agentic test instances). 
Aegis-Bench contains over 9K trajectories collected from six MASs and covers 14 failure modes, with each trajectory potentially involving multiple faulty agents and failure modes. Following its default setup, we use 600 trajectories for testing and split the remainder 8:2 into training and validation sets.

\paragraph{Baselines.}
We compare ASCon with several advanced prompt-based and learning-based failure attribution methods. For \textbf{root-fault attribution}, the prompt-based baselines include All-at-Once, Step-by-Step, and Binary Search~\citep{zhang2025who&when}, as well as SDBL (EASD)~\citep{sun2026scope}, all instantiated with GPT-4o-mini following their original settings.  
We additionally evaluate DeepSeek-V4-Pro and Gemini-3.1-Flash-lite with the All-at-Once prompt. The learning-based baselines include StepFinder~\citep{zhu2026stepfinder}, AgentTracer~\citep{zhang2026agentracer}, and Qwen-SFT, fine-tuned from Qwen2.5-14B.
For \textbf{failure-mode attribution} where related baselines remain limited, we evaluate the same commercial LLMs using the prompt template from Aegis~\citep{kong2026aegis}, and include Aegis-SFT (fine-tuned with LoRA), as the learning-based baseline. All prompt templates used in our evaluation are provided in Appendix~A.

\paragraph{Metrics.}
Following prior studies~\citep{kong2026aegis,zhu2026stepfinder}, we evaluate root-fault attribution using agent- and step-level accuracies, which measures whether the top-ranked prediction matches the ground-truth. For failure-mode attribution, we report Micro-F1(\(\mu\)F1) and Macro-F1(MF1). $\mu$F1 reflects overall performance across all instances, whereas MF1 computes the F1-score for each class independently and then takes the unweighted average.

\paragraph{Parameters.}
We set $L_s$ and $L_a$ to 2, the step and agent representation dimensions to 256, the attention dimension to 128, and the dropout rate to 0.1. Qwen3-Embedding-0.6B is used to encode steps and agent roles. All models are trained for 12 epochs with AdamW, using a learning rate of $5\times10^{-5}$ and a weight decay of $10^{-5}$. The ASCon model is implemented in PyTorch and runs on an Intel Core Ultra 7 265KF CPU with an NVIDIA GeForce RTX 5060 GPU.

\subsection{Main Results}

\subsubsection{Performance on root-fault attribution.}
Table~\ref{tab:root-fault} shows that ASCon achieves the best overall performance, with macro-accuracies of 81.58\% for faulty agent attribution and 63.90\% for faulty step attribution. Compared with StepFinder, the strongest baseline, ASCon improves the two metrics by 3.38\% and 7.57\%, respectively. 
Across different task domains, ASCon achieves the best results on code and agentic tasks and matches StepFinder on math, with particularly large step-level gains of 11.03\% on code and 11.66\% on agentic tasks.
The consistently lower step-level scores than agent-level scores indicate that locating the exact faulty action is more challenging than identifying the responsible agent. Across the baselines, the  agent-step macro-accuracy gap ranges from 21.87\% to 40.63\%, whereas ASCon reduces it to 17.68\%. This is mostly because ASCon incorporates direction-aware step context and agent-level behavioral context into each step representation, thereby providing stronger evidence for fine-grained fault localization.

\begin{table}[t]
\centering
\renewcommand{\arraystretch}{1}
\setlength{\tabcolsep}{2pt}
\fontsize{10pt}{10.4pt}\selectfont
\begin{tabular}{p{2.6cm}cccccc}
\toprule
\multirow{2}{*}{\makecell[c]{\textbf{Model}}}
& \multicolumn{2}{c}{\textbf{Pair}} 
& \multicolumn{2}{c}{\textbf{Agent}} 
& \multicolumn{2}{c}{\textbf{Error}}  \\
\cmidrule(lr){2-3} \cmidrule(lr){4-5} \cmidrule(lr){6-7}
& $\mu$F1 & MF1 & $\mu$F1 & MF1 & $\mu$F1 & MF1 \\
\midrule
DeepSeek-V4-Pro      & 10.70 & 4.70  & 73.40 & 46.78 & 22.04 & 16.19 \\
Gemini-3.1-Flash   & 12.67 & 5.97  & 71.76 & 43.65 & 23.34 & 17.11 \\
GPT-4o-mini            & 9.86  & 3.14  & 65.28 & 27.72 & 20.81 & 16.32 \\
Qwen3.5-Flash          & 11.49 & 4.70  & 67.69 & 40.50 & 23.61 & 16.35 \\
Aegis-SFT  & 17.85 &9.89 &\textbf{86.58} &68.21 &26.08&22.96\\
\midrule
\textbf{Our Method}               & \textbf{31.80} & \textbf{20.98} & 86.48 & \textbf{71.31} & \textbf{38.27} & \textbf{37.69}  \\
\bottomrule
\end{tabular}
\caption{Results for the failure-mode attribution task on the Aegis-Bench.
}
\label{tab:aegis_llm_fault_type}
\end{table}

\subsubsection{Performance on failure-mode attribution}
We evaluate failure-mode attribution at three levels:
 \textit{Agent}, which identifies faulty agents while ignoring failure modes; \textit{Error}, which identifies failure modes while ignoring agents; and \textit{Pair}, which requires the correct agent-error association.
 Table~\ref{tab:aegis_llm_fault_type} shows that ASCon achieves the best overall performance at all level attributions. Compared with Aegis-SFT, the strongest baseline, ASCon improves pair-level \(\mu\)F1 and MF1 by 13.95\% and 11.09\%, respectively, while increasing error-level \(\mu\)F1 and MF1 by 12.19\% and 14.73\%. At the agent level, ASCon improves MF1 by 3.10\% while maintaining comparable \(\mu\)F1, with a difference of only 0.10\%. These results indicate that ASCon is more effective at jointly identifying faulty agents and their associated error types, while the consistent MF1 gains suggest improved performance across less frequent classes rather than only on dominant categories.

\begin{table}[t]
\centering
\setlength{\tabcolsep}{0.1mm}
\fontsize{10pt}{9.5pt}\selectfont
\begin{tabular}{@{}l@{\hspace{-2mm}}ccc@{}c@{}}
 \hline
\toprule
 \multicolumn{5}{c}{\textbf{Root-fault attribution } }\\
\cmidrule(lr){1-5}
\textbf{Model} & \small \textbf{Agent $\mu$Acc} & \small \textbf{Agent MAcc}&   \small \textbf{Step $\mu$Acc} &  \small  \textbf{Step Acc}  \\
\addlinespace[1pt]
\multicolumn{5}{@{}l}{\textit{Lightweight models} ($<10$M)} \\
StepFinder & 42.93& 42.52 &19.02 &15.28 \\
ASCon
& 45.65 & 46.36
& 20.10 & 17.94 \\
 \cmidrule(lr){2-5}

\multicolumn{5}{@{}l}{\textit{LLM-based Models}  ($\geq$8B)} \\
DeepSeek-V4-Pro
& 58.70 & 55.42
& 22.28 & 17.20 \\
\quad\quad +ASCon &\textbf{67.39} &\textbf{66.89} & \textbf{43.48} &\textbf{40.59} \\ 
 \cmidrule(lr){2-5}
Gemini-3.1-Flash
& 53.80 & 49.99
& 22.83 & 18.06 \\
\quad\quad +ASCon &60.87 &57.94 &36.41 &31.24 \\

 \cmidrule(lr){2-5}
SDBL (EASD)
& 57.07 & 56.09
& 30.98 & 27.27 \\
\quad\quad +ASCon & 57.07 &60.28 & 34.24 &31.05 \\
 \cmidrule(lr){2-5}
 
Qwen-SFT & 50.00 & 49.07 & 27.17 & 25.42\\
\quad\quad +ASCon & 53.80 & 52.78 & 32.07& 29.93\\
 \cmidrule(lr){2-5}
 
AgentTracer
& 50.00 & 43.95
& 26.63 & 21.77 \\
\quad\quad +ASCon &55.43 &51.64 &30.43 &26.41 \\

 \hline
 \toprule
 \multicolumn{5}{c}{\textbf{Failure-mode attribution}} \\
\cmidrule(lr){1-5}
Model & \small Agent $\mu$F1 & \small Agent MF1& \small Error $\mu$F1& \small Error MF1  \\
\addlinespace[1pt]
ASCon & 44.76 & 26.43 & 7.37 & 5.95\\
 \cmidrule(lr){2-5}
 Deepseek-V4-pro & 48.10 & 29.60 & 12.64 & 7.57 \\
\quad\quad +ASCon & 51.14 & 38.23 & 13.82 & 7.76 \\
 \cmidrule(lr){2-5}
Gemini-3.1-Flash & 50.63 & 34.47 & 14.13 & 8.09 \\
\quad\quad +ASCon & 53.23 & 35.08 & \textbf{15.61} & \textbf{11.21} \\
 \cmidrule(lr){2-5}
GPT-4o-mini& 49.34 & 31.61 & 6.99& 5.16 \\
\quad\quad +ASCon & \textbf{53.90} & \textbf{38.74} & 10.33 & 8.71 \\
 \cmidrule(lr){2-5}
Qwen3.5-Flash & 46.04 & 28.97 & 9.22 & 6.46 \\
\quad\quad +ASCon & 52.52 & 31.85 & 11.89 & 9.34 \\
 \cmidrule(lr){2-5}
 Aegis-SFT & 31.78 &16.48 &5.39&2.85 \\
 \quad\quad +ASCon & 42.73 &27.09 &9.28 &6.43 \\
  \hline
   \toprule
\end{tabular}
\caption{Out-of-domain evaluation results on Who\&When dataset.``+ASCon'' denotes the use of ASCon predictions as auxiliary attribution evidence. $\mu$Acc and MAcc represent micro-accuracy and macro-accuracy, respectively.}
\label{tab:ood}
\end{table}

\begin{table*}[t]
\centering
\setlength{\tabcolsep}{2.0pt}
\renewcommand{\arraystretch}{1.08}
\small
\begin{tabular}{lcccccccccc}
\toprule
\multirow{3}{*}{\textbf{Setting}}
& \multicolumn{4}{c}{\textbf{Root-Fault Attribution}}
& \multicolumn{6}{c}{\textbf{Failure-Mode Attribution}} \\
\cmidrule(lr){2-5}\cmidrule(lr){6-11}

& \multicolumn{2}{c}{\textbf{Step}}
& \multicolumn{2}{c}{\textbf{Agent}}
& \multicolumn{2}{c}{\textbf{Agent}}
& \multicolumn{2}{c}{\textbf{Error}}
& \multicolumn{2}{c}{\textbf{Pair}} \\
\cmidrule(lr){2-3}
\cmidrule(lr){4-5}
\cmidrule(lr){6-7}
\cmidrule(lr){8-9}
\cmidrule(lr){10-11}

& $\mu$Acc & MAcc
& $\mu$Acc & MAcc
& $\mu$F1 & MF1
& $\mu$F1 & MF1
& $\mu$F1 & MF1 \\
\midrule
ASCon & 66.20$_{+0.00}$ & 
64.50$_{+0.00}$ & 
82.41$_{+0.00}$ & 
86.60$_{+0.00}$ & 
86.48$_{+0.00}$ & 
71.31$_{+0.00}$ & 38.27$_{+0.00}$ & 37.69$_{+0.00}$ & 31.80$_{+0.00}$ & 20.98$_{+0.00}$ \\
DGAT$\rightarrow$GAT & \underline{61.14}$_{-5.06}$ & 59.81$_{-4.69}$ & 79.37$_{-3.04}$ & \underline{82.24}$_{-4.36}$ & 83.10$_{-3.38}$ & 66.08$_{-5.23}$ & 33.67$_{-4.60}$ & 32.49$_{-5.20}$ & 26.06$_{-5.74}$ & 18.28$_{-2.70}$ \\
DGAT$\rightarrow$BiGRU & 64.94$_{-1.26}$ & \underline{58.57}$_{-5.93}$ & 81.27$_{-1.14}$ & 84.42$_{-2.18}$ & \underline{82.65}$_{-3.84}$ & \underline{62.17}$_{-9.14}$ & \underline{32.70}$_{-5.57}$ & \underline{31.98}$_{-5.71}$ & \underline{24.90}$_{-6.90}$ & \underline{15.87}$_{-5.11}$ \\
w/o MSTAA & 62.03$_{-4.17}$ & 61.68$_{-2.82}$ & \underline{76.84}$_{-5.57}$ & 82.87$_{-3.74}$ & 85.13$_{-1.35}$ & 68.88$_{-2.43}$ & 36.94$_{-1.33}$ & 35.70$_{-1.99}$ & 29.12$_{-2.68}$ & 18.54$_{-2.44}$ \\
MSTAA$\rightarrow$Mean & 61.39$_{-4.81}$ & 64.17$_{-0.33}$ & 80.76$_{-1.65}$ & 85.67$_{-0.93}$ & 85.98$_{-0.50}$ & 68.49$_{-2.82}$ & 34.79$_{-3.48}$ & 34.38$_{-3.31}$ & 27.82$_{-3.98}$ & 17.89$_{-3.09}$ \\
w/o ACSR & \underline{61.14}$_{-5.06}$ & 62.62$_{-1.88}$ & 82.03$_{-0.38}$ & 86.92$_{+0.32}$ & 85.69$_{-0.79}$ & 62.58$_{-8.73}$ & 36.73$_{-1.54}$ & 35.50$_{-2.19}$ & 29.16$_{-2.64}$ & 18.66$_{-2.32}$ \\

\bottomrule
\end{tabular}

\caption{Ablation results of ASCon. 
Subscripts denote changes relative to the full model. Underlined values indicate the largest degradation for each metric.
}
\label{tab:ablation}
\end{table*}

\subsection{Generalization to Out-of-Domain Scenarios}

To evaluate out-of-domain generalization, we test ASCon on Who\&When benchmark without further training.
Beyond default baselines, we incorporate ASCon predictions into several strong LLM-based attribution methods to examine whether its fault evidence can complement their reasoning under distribution shift. Specifically, for SDBL, we replace its original fault-range prediction with the candidates ranked highest by ASCon, selecting the top-3 agents and the top-5/10 steps for the Algorithm-Generated/Handcrafted subset, while retaining its subsequent reasoning process. 
For other LLM-based baselines, ASCon predictions are appended to the input. Root-fault attribution uses step and agent fault probabilities, while failure-mode attribution uses agent-level fault probabilities and top-five candidate fault types. Detailed formats, settings, and results are provided in Appendix~B.

As shown in Table~\ref{tab:ood}, ASCon outperforms StepFinder on all agent- and step-level metrics, indicating stronger out-of-domain transfer among lightweight attribution models. Some LLM-based methods achieve higher performance, which is expected given their substantially larger model capacity and broad pretrained knowledge. Nevertheless, incorporating ASCon predictions further improves these methods on all granularity metrics. These results show that ASCon captures transferable fine-grained fault evidence that effectively complements LLM-based reasoning in unseen scenarios.

\subsection{Ablations}

To assess the contribution of each component in ASCon, we construct the following five variants. \textit{DGAT$\rightarrow$GAT} and \textit{DGAT$\rightarrow$BiGRU} replace DGAT with a standard GAT and a bidirectional GRU, respectively, to evaluate the necessity of direction-aware relational modeling. \textit{w/o MSTAA} and \textit{MSTAA$\rightarrow$Mean} remove masked step-to-agent attention and replace its learned weighting with mean pooling over agent-owned steps, respectively, to examine the importance of selective behavioral aggregation. Finally, \textit{w/o ACSR} removes the agent-conditioned step representation to assess the contribution of agent context to step-level attribution.

As shown in Table~\ref{tab:ablation}, all variants exhibit overall performance degradation, confirming that the three components provide complementary attribution evidence. Replacing DGAT consistently reduces performance, while BiGRU causes the largest degradation on most failure-mode metrics, including a 9.14\% drop in Agent MF1 and a 6.90\% drop in Pair MF1. This suggests that sequential encoding alone is insufficient, and that explicitly modeling directional relations among steps provides more discriminative structural evidence. 
The degradation of the MSTAA variants shows that learned step-to-agent attention is more effective than simple averaging, as different steps contribute unequally and irrelevant or propagated behaviors should be suppressed.
 Removing ACSR substantially reduces root-fault Step micro-accuracy by 5.06\% and consistently harms Error and Pair attribution, showing that conditioning individual steps on their responsible agents improves fine-grained agent--error association.
Overall, the ablation results support the complementary roles of directional step modeling, selective step-to-agent aggregation, and agent-conditioned step refinement.

\subsection{Can ASCon Capture Potential Dependencies?}
\begin{figure}[htbp]
    \centering
    \includegraphics[width=1.\linewidth]{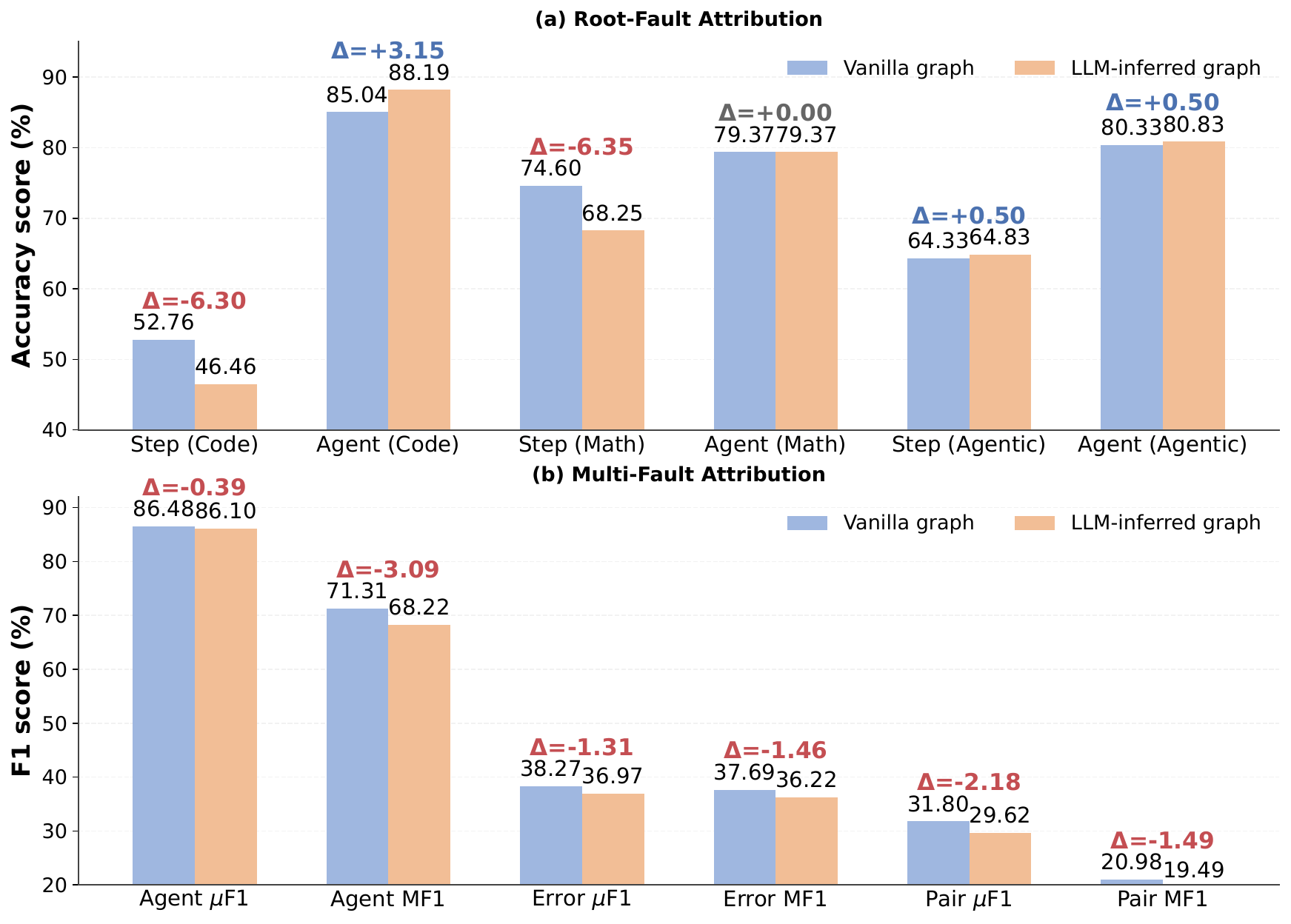}
    \caption{Performance comparison between vanilla graphs and LLM-inferred graphs. 
    }
    \label{fig:llm_graph_comparison}
\end{figure}
To assess whether ASCon can implicitly learn useful inter-step and inter-agent dependencies from the vanilla graph, we replace its original edges with dependencies inferred by Deepseek-V4-pro with the prompt in Appendeix A.7.
As shown in Figure~\ref{fig:llm_graph_comparison}, the LLM-inferred graph provides no consistent advantage. In root-fault attribution, it improves code agent accuracy by 3.15\% and yields marginal gains on agentic tasks, but reduces step accuracy by 6.30\% on code and 6.35\% on math. In failure-mode attribution, the two graph constructions perform comparably, with the LLM-inferred graph only slightly lower. 
 These results suggest that ASCon can already capture attribution-relevant dependencies through supervised representation learning and directional graph aggregation. In contrast, LLM-inferred edges may introduce noisy or overly broad connections, which are particularly harmful to fine-grained step localization and failure-mode discrimination. Therefore, the simpler vanilla graph in ASCon is both sufficient and more reliable.

\section{Conclusion}
This paper presented ASCon, a direction-aware reciprocal agent-step contextualization model for MAS failure attribution. By jointly modeling directional action context, agent behavioral histories, and inter-agent interactions, ASCon learns contextualized representations that support faulty-agent, faulty-step, and failure-mode attribution. Experiments demonstrate ASCon made consistent improvements over strong baselines, and can also enhance existing attribution methods in out-of-domain scenarios. These findings highlight ASCon can learn effective agent and step contextualization representations for fine-grained and transferable MAS failure diagnosis.

\bibliography{References}
\bibliographystyle{plainnat}

\newpage
\clearpage
\appendix
\onecolumn

\begin{center}
    {\LARGE\bfseries Appendix}
\end{center}

\section{A. Prompt Templates Used in This Paper}
\label{ap-1}
\vspace{0.3cm}

\subsubsection{A.1 All-at-Once Attribution Prompt Template for Root-Fault Attribution}
\mbox{}\par
\vspace{0.3cm}
This prompt template comes frome the All-at-Once baseline ~\citep{zhang2025who&when}, and is also used by Qwen-SFT and the LLM-based baselines for root-fault attribution.
\begin{promptbox}
\small
You are an AI assistant tasked with analyzing a multi-agent conversation
history when solving a real-world problem.

The problem is: \textcolor{black}{\{problem\}}.

Identify which agent made an error, at which step, and explain the reason
for the error.

Here is the conversation:
\textcolor{black}{\{chat\_content\}}

Based on this conversation, please predict the following:

\begin{enumerate}
    \item The name of the agent who made a mistake and should be directly
    responsible for the incorrect solution to the real-world problem.
    If no agent makes an obvious mistake, select the single most likely
    responsible agent. Directly output the name of the agent.

    \item The step at which the responsible agent first made a mistake.
    For example, consider the following conversation:
    \[
    \{
    \text{``agent a'': ``xx'', }
    \text{``agent b'': ``xxxx'', }
    \text{``agent c'': ``xxxxx'', }
    \text{``agent a'': ``xxxxxxx''}
    \}.
    \]
    Each entry represents one step in which an agent provides an output.
    If the mistake occurs in agent c's output, the step number is 2.
    If it occurs in the second output of agent a, the step number is 3.

    \item The reason for your prediction.
\end{enumerate}

Please answer in valid JSON format as follows:

\begin{Verbatim}[
    fontsize=\footnotesize,
    breaklines=true,
    breakanywhere=true,
    breaksymbolleft={},
    breaksymbolright={},
    breakindent=0pt
]
{
  "agent_name": "Your predicted root-cause faulty agent name",
  "step_number": 1,
  "reason_for_mistake": "Your reason"
}
\end{Verbatim}

\end{promptbox}

\subsubsection{A.2 Prompt of AgentTracer}
\mbox{}\par\vspace{0.4em}

\begin{promptbox}
\small

You are AgentTracer, a failure attribution model for LLM-based agentic systems.

Given a failed multi-agent trajectory, identify the earliest decisive error: the first step whose correction would be sufficient to avoid the final failure.

Return exactly the following format:

\begin{Verbatim}[
    fontsize=\footnotesize,
    breaklines=true,
    breakanywhere=true,
    breaksymbolleft={},
    breaksymbolright={},
    breakindent=0pt
]
{
<think>brief evidence-based reasoning</think>

<answer>agent_name|step_number</answer>}
\end{Verbatim}

The \texttt{agent\_name} must match one of the agents in the trajectory, and \texttt{step\_number} must be an integer step ID.

\end{promptbox}

\subsubsection{A.3 Prompt of ASCon-Enhanced SDBL Method}
\mbox{}\par\vspace{0.4em}

\begin{promptbox}
\small

You are an AI assistant tasked with analyzing a multi-agent conversation history when solving a real-world problem.

The problem is: \textcolor{black}{\{problem\}}.

Identify which agent made an error, at which step, and explain the reason for the error.

Here is the conversation:

\textcolor{black}{\{chat\_content\}}

\textbf{The following agents and steps are flagged for special attention. These are the candidates identified by our automated procedure as the most likely fault locations. Please focus on these high-confidence candidates and prioritize them when locating the root cause:}

\textbf{\textcolor{black}{\{reference\_content\}}}

Based on this conversation, please predict the following:

\begin{enumerate}
    \item The name of the agent directly responsible for the incorrect solution. If no agent makes an obvious mistake, select the single most likely responsible agent and output its name.

    \item The step at which the responsible agent first made a mistake. For example, in the conversation
    \[
    \{
    \text{``agent a'': ``xx'', }
    \text{``agent b'': ``xxxx'', }
    \text{``agent c'': ``xxxxx'', }
    \text{``agent a'': ``xxxxxxx''}
    \},
    \]
    each entry represents one step. If the mistake occurs in agent c's output, the step number is 2. If it occurs in the second output of agent a, the step number is 3.

    \item The reason for your prediction.
\end{enumerate}

Please answer in valid JSON format as follows:

\begin{Verbatim}[
  fontsize=\footnotesize,
  breaklines=true,
  breakanywhere=true,
  breaksymbolleft={},
  breaksymbolright={},
  breakindent=0pt
]
{
  "agent_name": "Your predicted root-cause faulty agent name",
  "step_number": 1,
  "reason_for_mistake": "Your reason"
}
\end{Verbatim}

\end{promptbox}

\subsubsection{A.4 Prompt of ASCon-Enhanced LLM-Based Attribution Methods}
\mbox{}\par\vspace{0.4em}

\begin{promptbox}
\small

You are an AI assistant tasked with analyzing a multi-agent conversation history when solving a real-world problem.

The problem is: \textcolor{black}{\{problem\}}.

Identify which agent made an error, at which step, and explain the reason for the error.

Here is the conversation:

\textcolor{black}{\{chat\_content\}}

Based on this conversation, please predict the following:

\begin{enumerate}
    \item \textbf{Each step is associated with an auxiliary fault probability. First examine the high-probability steps and their neighboring context to determine whether they contain the root-cause mistake.}

    \item The name of the agent directly responsible for the incorrect solution. If no agent makes an obvious mistake, select the single most likely responsible agent and output its name.

    \item The step at which the responsible agent first made a mistake. For example, in the conversation
    \[
    \{
    \text{``agent a'': ``xx'', }
    \text{``agent b'': ``xxxx'', }
    \text{``agent c'': ``xxxxx'', }
    \text{``agent a'': ``xxxxxxx''}
    \},
    \]
    each entry represents one step. If the mistake occurs in agent c's output, the step number is 2. If it occurs in the second output of agent a, the step number is 3.

    \item The reason for your prediction.
\end{enumerate}

Please answer in valid JSON format as follows:

\begin{Verbatim}[
  fontsize=\footnotesize,
  breaklines=true,
  breakanywhere=true,
  breaksymbolleft={},
  breaksymbolright={},
  breakindent=0pt
]
{
  "agent_name": "Your predicted root-cause faulty agent name",
  "step_number": 1,
  "reason_for_mistake": "Your reason"
}
\end{Verbatim}

\end{promptbox}

\subsubsection{A.5 Prompt Template for Failure-Mode Attribution}
\mbox{}\par\vspace{0.4em}

\begin{promptbox}
\small

\#\# ROLE AND GOAL
You are a meticulous Multi-Agent System (MAS) Quality Assurance analyst. Your sole purpose is to analyze conversation logs to identify and categorize agent errors based on a strict set of definitions.
\\
\\
\#\# ERROR DEFINITIONS WITH EXAMPLES

You MUST use the exact error codes provided below.
\#\#\# Functional Mistakes (FM-1.x - Task Execution Errors):

- FM-1.1: **Task specification deviation** - Agent deviates from specified task requirements (e.g., was asked to write code in Python, but used JavaScript).

- FM-1.2: **Role specification deviation** - Agent acts outside its designated role (e.g., a 'CodeWriter' agent starts criticizing other agents' work, which is the 'Critic's' role).

- FM-1.3: **Add redundant steps** - Agent adds unnecessary or duplicate steps (e.g., imports a library that was already imported in a previous step).

- FM-1.4: **Remove conversation history** - Agent ignores or removes important context from previous turns (e.g., ignores a user's correction from the previous message).

- FM-1.5: **Remove termination conditions** - Agent fails to define proper stopping criteria, leading to loops or unfinished tasks (e.g., writes a recursive function with no base case).
\#\#\# Functional Mistakes (FM-2.x - Communication \& Coordination Errors):

- FM-2.1: **Repeat handled tasks** - Agent redundantly handles already completed tasks (e.g., re-writes a piece of code that was already finalized and approved).

- FM-2.2: **Make request ambiguous** - Agent provides unclear or confusing instructions to other agents (e.g., asks another agent to "handle the data" without specifying how).

- FM-2.3: **Deviate from main goal** - Agent pursues objectives unrelated to the main task (e.g., starts discussing the history of programming languages in the middle of a coding task).

- FM-2.4: **Hide important information** - Agent withholds crucial information needed by other agents (e.g., knows a library has a bug but doesn't mention it).

- FM-2.5: **Ignore other agents** - Agent fails to consider input, corrections, or questions from other agents.

- FM-2.6: **Inconsistent reasoning** - Agent's logic contradicts its own previous statements (e.g., in step 2 agent says 'option A is best', but in step 4 says 'option A is a bad choice' without new information).
\#\#\# Functional Mistakes (FM-3.x - Quality \& Verification Errors):

- FM-3.1: **Premature termination** - Agent stops or declares the task complete before all requirements are met.

- FM-3.2: **Remove verification steps** - Agent skips necessary validation or testing steps (e.g., writes code but doesn't write any unit tests for it).

- FM-3.3: **Incorrect verification** - Agent performs flawed or wrong verification (e.g., writes a test that doesn't actually check for the correct condition).
\\
\\
\#\# ANALYSIS WORKFLOW

Please follow these steps carefully:

\#\#\# Step 1: Agent Summary

First, analyze and summarize what each agent has done throughout the conversation:

- List each agent that appears in the conversation
- For each agent, summarize their main actions, decisions, and contributions

- Note any patterns or recurring behaviors

\#\#\# Step 2: Error Analysis

For each agent identified in Step 1:

- Carefully examine their actions against each error definition

- Look for violations of task requirements, role boundaries, communication issues, or quality problems

- Note any potential errors with specific reasoning

\#\#\# Step 3: Final Judgment

Based on your analysis in Steps 1 and 2:

- Determine which agents (if any) committed errors

- Assign the appropriate error code(s) to each faulty agent

- Ensure agent names match exactly as they appear in the conversation log
\\
\\
\#\# REQUIRED OUTPUT FORMAT

Your response must contain:

1. **Agent Summary**: A brief analysis of what each agent did
2. **Error Analysis**: Your reasoning for identifying errors
3. **Final Answer**: A valid JSON object with your conclusions

**JSON Format:**
{{"faulty\_agents": [{{"agent\_name": "XXX", "error\_type": "FM-X.X"}}]}}

**Examples:**
- Multiple Errors: {{"faulty\_agents": [{{"agent\_name": "XXX1", "error\_type": "FM-1.1"}}, {{"agent\_name": "XXX2", "error\_type": "FM-3.2"}}, {{"agent\_name": "XXX3", "error\_type": "FM-2.5"}}]}}
- No Errors: {{"faulty\_agents": []}}

**Important:** Make sure the agent names you output exactly match those in the conversation log. Do not fabricate names.
\\
\\
\#\# CONVERSATION TO ANALYZE:
\textcolor{black}{\{conversation\_text\}}
\\
\\
\#\# YOUR ANALYSIS:

\end{promptbox}

\subsubsection{A.6 Prompt Template for ASCon-enhaced Failure-Mode Attribution}
\mbox{}\par\vspace{0.4em}

\begin{promptbox}
\small

\#\# ROLE AND GOAL
You are a meticulous Multi-Agent System (MAS) Quality Assurance analyst. Your sole purpose is to analyze conversation logs to identify and categorize agent errors based on a strict set of definitions.
\\
\\
\#\# ERROR DEFINITIONS WITH EXAMPLES

You MUST use the exact error codes provided below.
\#\#\# Functional Mistakes (FM-1.x - Task Execution Errors):

- FM-1.1: **Task specification deviation** - Agent deviates from specified task requirements (e.g., was asked to write code in Python, but used JavaScript).

- FM-1.2: **Role specification deviation** - Agent acts outside its designated role (e.g., a 'CodeWriter' agent starts criticizing other agents' work, which is the 'Critic's' role).

- FM-1.3: **Add redundant steps** - Agent adds unnecessary or duplicate steps (e.g., imports a library that was already imported in a previous step).

- FM-1.4: **Remove conversation history** - Agent ignores or removes important context from previous turns (e.g., ignores a user's correction from the previous message).

- FM-1.5: **Remove termination conditions** - Agent fails to define proper stopping criteria, leading to loops or unfinished tasks (e.g., writes a recursive function with no base case).
\#\#\# Functional Mistakes (FM-2.x - Communication \& Coordination Errors):

- FM-2.1: **Repeat handled tasks** - Agent redundantly handles already completed tasks (e.g., re-writes a piece of code that was already finalized and approved).

- FM-2.2: **Make request ambiguous** - Agent provides unclear or confusing instructions to other agents (e.g., asks another agent to "handle the data" without specifying how).

- FM-2.3: **Deviate from main goal** - Agent pursues objectives unrelated to the main task (e.g., starts discussing the history of programming languages in the middle of a coding task).

- FM-2.4: **Hide important information** - Agent withholds crucial information needed by other agents (e.g., knows a library has a bug but doesn't mention it).

- FM-2.5: **Ignore other agents** - Agent fails to consider input, corrections, or questions from other agents.

- FM-2.6: **Inconsistent reasoning** - Agent's logic contradicts its own previous statements (e.g., in step 2 agent says 'option A is best', but in step 4 says 'option A is a bad choice' without new information).
\#\#\# Functional Mistakes (FM-3.x - Quality \& Verification Errors):

- FM-3.1: **Premature termination** - Agent stops or declares the task complete before all requirements are met.

- FM-3.2: **Remove verification steps** - Agent skips necessary validation or testing steps (e.g., writes code but doesn't write any unit tests for it).

- FM-3.3: **Incorrect verification** - Agent performs flawed or wrong verification (e.g., writes a test that doesn't actually check for the correct condition).
\\
\\
\#\# ANALYSIS WORKFLOW

Please follow these steps carefully:

\#\#\# Step 1: Agent Summary

First, analyze and summarize what each agent has done throughout the conversation:

- List each agent that appears in the conversation
- For each agent, summarize their main actions, decisions, and contributions

- Note any patterns or recurring behaviors

\#\#\# Step 2: Error Analysis

For each agent identified in Step 1:

- Carefully examine their actions against each error definition

- Look for violations of task requirements, role boundaries, communication issues, or quality problems

- Note any potential errors with specific reasoning

\#\#\# Step 3: Final Judgment

Based on your analysis in Steps 1 and 2:

- Determine which agents (if any) committed errors

- Assign the appropriate error code(s) to each faulty agent

- Ensure agent names match exactly as they appear in the conversation log
\\
\\
\textbf{\#\#\# AUXILIARY AGENT FAULT PROFILE
\\
Each conversation will be followed by an auxiliary fault profile for every agent, including:
\\
- auxiliary fault probability: the probability that the agent is faulty in this task;
\\
- top 5 likely fault types: the five most likely fault types predicted for that agent. 
\\
Please use this auxiliary information to assist your judgment of faulty agents and fault types.
}
\\
\\
\#\# REQUIRED OUTPUT FORMAT

Your response must contain:

1. **Agent Summary**: A brief analysis of what each agent did
2. **Error Analysis**: Your reasoning for identifying errors
3. **Final Answer**: A valid JSON object with your conclusions

**JSON Format:**
{{"faulty\_agents": [{{"agent\_name": "XXX", "error\_type": "FM-X.X"}}]}}

**Examples:**
- Multiple Errors: {{"faulty\_agents": [{{"agent\_name": "XXX1", "error\_type": "FM-1.1"}}, {{"agent\_name": "XXX2", "error\_type": "FM-3.2"}}, {{"agent\_name": "XXX3", "error\_type": "FM-2.5"}}]}}
- No Errors: {{"faulty\_agents": []}}

**Important:** Make sure the agent names you output exactly match those in the conversation log. Do not fabricate names.
\\
\\
\#\# CONVERSATION TO ANALYZE:
\textcolor{black}{\{conversation\_text\}}
\\
\\
\#\# YOUR ANALYSIS:

\end{promptbox}

\subsubsection{A.7 Prompt of LLM-Based Agent/Step Graph Inference}
\mbox{}\par\vspace{0.4em}

\begin{promptbox}
\small

You are an expert in Multi-Agent System behavior analysis. I will provide you
with a trajectory of agent behaviors consisting of multiple numbered steps.
Each step contains information such as the executing agent, actions, and
dialogues. You need to complete two tasks and output the results strictly in
the specified format.

\medskip
\textbf{Task 1: Construct the Step Dependency Graph}

Nodes are step numbers, i.e., Step 1, Step 2, \ldots. For any two steps
$S_i$ and $S_j$, typically with $i<j$, establish a directed edge
$S_i \rightarrow S_j$ if any of the following conditions is satisfied:

\begin{enumerate}
    \item \textbf{Content Dependency:}
    The output of $S_j$ retains, integrates, references, or is directly
    derived from the concrete output of $S_i$. This includes data,
    instructions, assumptions, intermediate conclusions, code snippets,
    or task descriptions.

    \item \textbf{Goal/Context Dependency:}
    Step $S_i$ establishes the problem definition, task goal, execution
    context, role assignment, or operational constraints used by $S_j$.

    \item \textbf{Feedback Dependency:}
    Step $S_j$ explicitly responds to, evaluates, or revises the output
    of step $S_i$.
\end{enumerate}

\textbf{Important notes for Task 1:}

\begin{itemize}
    \item Add only one edge when the same pair satisfies multiple conditions.
    \item A step may depend on multiple preceding steps.
    \item Do not omit dependencies originating from initialization,
    problem-definition, or task-assignment steps.
    \item Use the step numbers exactly as they appear in the trajectory.
\end{itemize}

\textbf{Task 2: Construct the Agent Interaction Graph}

Nodes are agent names. Add a directed edge between two agents when:

\begin{itemize}
    \item information, messages, tasks, requests, feedback, or intermediate
    conclusions are transmitted between them;
    \item one message or task explicitly targets multiple agents, in which
    case an edge is added from the sender to each recipient;
    \item a step dependency $S_i \rightarrow S_j$ exists, where $S_i$ is
    performed by agent A and $S_j$ by agent B, with $A\neq B$.
\end{itemize}

Do not add duplicate edges.

\medskip
\textbf{Output Format}

Output only a valid JSON object. Agent names must be copied from the
\texttt{name} attribute of the corresponding history steps.

\begin{Verbatim}[
fontsize=\footnotesize,
breaklines=true,
breakanywhere=true,
breaksymbolleft={},
breaksymbolright={}
]
{
  "agent_interactions": [
    {
      "from": "<Agent name>",
      "to": "<Agent name>"
    }
  ],
  "step_causalities": [
    {
      "from": <step_number>,
      "to": <step_number>
    }
  ]
}
\end{Verbatim}

Here is the agent behavior trace to be analyzed:

\texttt{\{trajectory\}}

\end{promptbox}

\subsubsection{A.8 Real-Time Failure Detection Prompt}
\mbox{}\par\vspace{0.4em}

\begin{promptbox}
\small

You are a real-time failure detector for a multi-agent system. We will provide you with the original user task and the execution trajectory observed so far. Your task is to determine, based on the observed agent actions and intermediate states, whether a failure has occurred.

A failure occurs when an observed action or intermediate result is inconsistent with the user task, established task constraints, valid prior information, or the requirements of the current execution step.

Do not report a failure solely because the task is unfinished or the trajectory contains exploration, uncertainty, or a later correction.

Return only:
\begin{Verbatim}[
fontsize=\footnotesize,
breaklines=true,
breakanywhere=true,
breaksymbolleft={},
breaksymbolright={}
]
{
{
  "fault\_detected": true or false,
  "fault\_step": the earliest identifiable faulty step ID or null,
  "fault\_agent": the responsible agent or null,
  "reason": "A brief explanation based on the observed trajectory."
}
}
\end{Verbatim}
Here is the user task and the execution trajectory.
[USER TASK]
{task}

[OBSERVED TRAJECTORY]
{trajectory}

\end{promptbox}


\section{B. Detailed Generalization Performance Comparison Results}
\begin{table*}[htbp]
\centering
\setlength{\tabcolsep}{1.5pt}

\begin{tabular}{l*{8}{l}}
\toprule
\multirow{2}{*}{\textbf{Model}}
& \multicolumn{2}{c}{\textbf{HC}}
& \multicolumn{2}{c}{\textbf{AG}}
& \multicolumn{2}{c}{\textbf{Micro-Accuracy}}
& \multicolumn{2}{c}{\textbf{Macro-Accuracy}} \\
\cmidrule(lr){2-3}
\cmidrule(lr){4-5}
\cmidrule(lr){6-7}
\cmidrule(lr){8-9}

& \textbf{Agent} & \textbf{Step}
& \textbf{Agent} & \textbf{Step}
& \textbf{Agent} & \textbf{Step}
& \textbf{Agent} & \textbf{Step} \\
\midrule

\multicolumn{9}{@{}l}{\textit{Lightweight models} ($<10$M)} \\
\addlinespace[1pt]

StepFinder
& 41.38 & 5.17
& 43.65 & 25.40
& 42.93 & 19.02
& 42.52 & 15.28 \\

ASCon
& 48.28 & 12.07
& 44.44 & 23.81
& 45.65 & 20.11
& 46.36 & 17.94 \\

\cmidrule(lr){1-9}

\multicolumn{9}{@{}l}{\textit{LLM-based models} ($\geq 8$B)} \\
\addlinespace[1pt]

Gemini-3.1-Flash
& 39.66 & 5.17
& 60.32 & 30.95
& 53.80 & 22.83
& 49.99 & 18.06 \\

GPT-4o-mini
& 58.62 & 6.90
& 53.97 & 11.90
& 55.44 & 10.32
& 56.30 & 9.40 \\

Step-by-Step
& 31.03 & 12.07
& 24.60 & 14.29
& 26.63 & 13.59
& 27.82 & 13.18 \\

Binary Search
& 20.17 & 10.17
& 27.78 & 10.32
& 25.38 & 10.27
& 23.98 & 10.25 \\

\cmidrule(lr){1-9}

DeepSeek-V4-Pro
& 46.55 & 3.45
& 64.29 & 30.95
& 58.70 & 22.28
& 55.42 & 17.20 \\

\quad\quad +ASCon
& 65.52$_{+18.97}$
& \textbf{32.76}$_{+29.31}$
& \textbf{68.25}$_{+3.96}$
& \textbf{48.41}$_{+17.46}$
& \textbf{67.39}$_{+8.69}$
& \textbf{43.48}$_{+21.20}$
& \textbf{66.89}$_{+11.47}$
& \textbf{40.59}$_{+23.39}$ \\

\cmidrule(lr){2-9}

Gemini-3.1-Flash
& 39.66 & 5.17
& 60.32 & 13.49
& 53.80 & 10.87
& 49.99 & 9.33 \\

\quad\quad +ASCon
& 50.00$_{+10.34}$
& 17.24$_{+12.07}$
& 65.87$_{+5.55}$
& 45.24$_{+31.75}$
& 60.87$_{+7.07}$
& 36.41$_{+25.54}$
& 57.94$_{+7.95}$
& 31.24$_{+21.91}$ \\

\cmidrule(lr){2-9}

SDBL (EASD)
& 53.45 & 17.24
& 58.73 & 37.30
& 57.07 & 30.98
& 56.09 & 27.27 \\

\quad\quad +ASCon
& \textbf{68.97}$_{+15.52}$
& 22.41$_{+5.17}$
& 51.59$_{-7.14}$
& 39.68$_{+2.38}$
& 57.07$_{+0.00}$
& 34.24$_{+3.26}$
& 60.28$_{+4.19}$
& 31.05$_{+3.78}$ \\

\cmidrule(lr){2-9}

Qwen-SFT
& 46.55 & 20.69
& 51.59 & 30.16
& 50.00 & 27.17
& 49.07 & 25.02 \\

\quad\quad +ASCon
& 50.00$_{+3.45}$
& 24.14$_{+3.45}$
& 55.56$_{+3.97}$
& 35.71$_{+5.55}$
& 53.80$_{+3.80}$
& 32.07$_{+4.90}$
& 52.78$_{+3.71}$
& 29.93$_{+4.91}$ \\

\cmidrule(lr){2-9}

AgentTracer
& 27.59 & 8.62
& 60.32 & 34.92
& 50.00 & 26.63
& 43.95 & 21.77 \\

\quad\quad +ASCon
& 41.38$_{+13.79}$
& 15.52$_{+6.90}$
& 61.90$_{+1.58}$
& 37.30$_{+2.38}$
& 55.43$_{+5.43}$
& 30.43$_{+3.80}$
& 51.64$_{+7.69}$
& 26.41$_{+4.64}$ \\

\bottomrule
\end{tabular}

\caption{Detailed results on the out-of-domain root-fault attribution task. Micro-accuracy is weighted by the HC and AG test-set sizes, whereas macro-accuracy assigns equal weight to both test sets. Subscripts in the ``+ASCon'' rows denote absolute percentage-point changes from the corresponding base methods. The best result in each column is shown in \textbf{bold}.}
\label{tab:detailed-ood1}
\end{table*}
\paragraph{Out-of-Domain Evaluation Setup.}
We evaluate the out-of-domain generalization of ASCon on the Who\&When benchmark without any further training or parameter updates. The root-fault attribution test set follows the original Who\&When dataset~\citep{zhang2025who&when}. The failure-mode attribution test set is adopted from ~\citep{kong2026aegis}, which extends the same Who\&When trajectories with additional failure-type annotations. 
In addition to evaluating ASCon directly, we incorporate its predictions into several strong LLM-based attribution methods to examine whether the learned fault evidence can complement LLM reasoning under distribution shift.
For root-fault attribution, we replace SDBL's original first-stage fault-range prediction with the candidates ranked highest by ASCon, while retaining its subsequent reasoning procedure. Specifically, we select the top-3 agents together with the top-5 and top-10 steps for the Algorithm-Generated (AG) and Handcrafted (HC) subsets, respectively. For the remaining LLM-based baselines, the predicted fault probability predicted by ASCon is attached to each step using the following format to enhance the root-fault attribution task:
\textit{\{step\_id: xxx, content: xxx, fault\_probability: xxx\}}. The augmented trajectory is then provided to the prompt templates in Appendices~A.3 and~A.4.

For failure-mode attribution, ASCon evidence is appended after the complete conversation history. Each agent is associated with its predicted fault probability and five most likely failure modes using the following format:
\textit{\{agent\_name: xxx, fault\_probability: xxx, top5\_likely\_failure\_modes: xxx\}}. The resulting input is evaluated using the prompt template in Appendix~A.6.

\paragraph{Out-of-Domain Results.}
Tables~\ref{tab:detailed-ood1} and~\ref{tab:detailed-ood2} report the out-of-domain results for root-fault and failure-mode attribution. On root-fault attribution, ASCon consistently outperforms StepFinder and substantially improves most enhanced baselines, especially for step localization. With ASCon assistance, DeepSeek-V4-Pro gains 21.20 and 23.39 percentage points in step-level micro- and macro-accuracy, respectively, and achieves the best results on seven of the eight metrics. These gains indicate that ASCon provides transferable fault evidence that helps LLMs identify where failures emerge under distribution shift.

For failure-mode attribution, LLMs assisted by ASCon achieve the best agent- and error-level results. Specifically, GPT-4o-mini with ASCon obtains the highest agent-level scores, while Gemini-3.1-Flash with ASCon achieves the best error-level performance. Aegis-SFT also receives the largest agent-level gains, improving by 10.95 and 10.61 percentage points in micro- and macro-F1. Overall, ASCon generalizes beyond its training distribution and effectively complements existing attribution methods.

\begin{table}[t]
\centering
\setlength{\tabcolsep}{2.2pt}
\begin{tabular}{l*{6}{l}}
\toprule
\textbf{Model}
& \textbf{Agent $\mu$F1}
& \textbf{Agent MF1}
& \textbf{Error $\mu$F1}
& \textbf{Error MF1}
& \textbf{Pair $\mu$F1}
& \textbf{Pair MF1} \\
\midrule

\multicolumn{7}{@{}l}{\textit{Lightweight models} ($<10$M)} \\
\addlinespace[1pt]

ASCon
& 44.76
& 26.43
& 7.37
& 5.95
& 1.87
& 1.12 \\

\cmidrule(lr){1-7}

\multicolumn{7}{@{}l}{\textit{LLM-based models} ($\geq 8$B)} \\
\addlinespace[1pt]

DeepSeek-V4-Pro
& 48.10
& 29.60
& 12.64
& 7.57
& 7.05
& 3.16 \\

\quad\quad +ASCon
& 51.14$_{+3.04}$
& 38.23$_{+8.63}$
& 13.82$_{+1.18}$
& 7.76$_{+0.19}$
& 8.24$_{+1.19}$
& \textbf{4.85}$_{+1.69}$ \\

\cmidrule(lr){2-7}

Gemini-3.1-Flash
& 50.63
& 34.47
& 14.13
& 8.09
& \textbf{8.41}
& \textbf{4.85} \\

\quad\quad +ASCon
& 53.23$_{+2.60}$
& 35.08$_{+0.61}$
& \textbf{15.61}$_{+1.48}$
& \textbf{11.21}$_{+3.12}$
& 7.64$_{-0.77}$
& 3.15$_{-1.70}$ \\

\cmidrule(lr){2-7}

GPT-4o-mini
& 49.34
& 31.61
& 6.99
& 5.16
& 3.40
& 1.70 \\

\quad\quad +ASCon
& \textbf{53.90}$_{+4.56}$
& \textbf{38.74}$_{+7.13}$
& 10.33$_{+3.34}$
& 8.71$_{+3.55}$
& 4.19$_{+0.79}$
& 2.51$_{+0.81}$ \\

\cmidrule(lr){2-7}

Qwen3.5-Flash
& 46.04
& 28.97
& 9.22
& 6.46
& 6.42
& 3.06 \\

\quad\quad +ASCon
& 52.52$_{+6.48}$
& 31.85$_{+2.88}$
& 11.89$_{+2.67}$
& 9.34$_{+2.88}$
& 6.69$_{+0.27}$
& 3.23$_{+0.17}$ \\

\cmidrule(lr){2-7}

Aegis-SFT
& 31.78
& 16.48
& 5.39
& 2.85
& 2.34
& 1.19 \\

\quad\quad +ASCon
& 42.73$_{+10.95}$
& 27.09$_{+10.61}$
& 9.28$_{+3.89}$
& 6.43$_{+3.58}$
& 3.96$_{+1.62}$
& 1.98$_{+0.79}$ \\

\bottomrule
\end{tabular}

\caption{Detailed results on the out-of-domain failure-mode attribution task.}
\label{tab:detailed-ood2}
\end{table}

\section{C. Can ASCon Perform Real-time Fault Detection?}
\begin{figure}[htbp]
    \centering
    \includegraphics[width=.5 \linewidth]{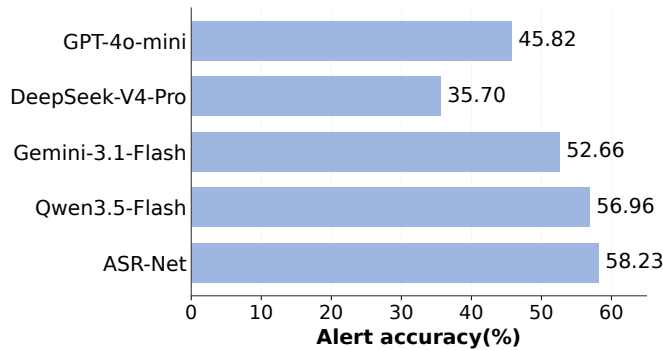}
    \caption{Performance comparison on the alert accuracy (\%).}
    \label{fig:realtime}
\end{figure}

To evaluate fault detection capabilities in an online setting, we construct a real-time detection experiment on the TracerTraj test set. Each trajectory is truncated at its first annotated faulty step, and only the user query and the interaction history observed up to that point are provided as input. We compare ASCon with four LLM-based reasoning baselines, including GPT-4o-mini, DeepSeek-V4-Pro, Gemini-3.1-Flash, and Qwen3.5-Flash, using the prompt provided in Appendix~A.8. A detection is considered as a correct alert only when the predicted faulty position exactly matches the ground-truth faulty position; both premature detection and missed detection are counted as errors.
Figure~\ref{fig:realtime} shows that ASCon achieves the highest correct alert rate of 58.23\%, outperforming Qwen3.5-Flash by 1.27\% and Gemini-3.1-Flash by 5.57\%. This result indicates that ASCon can identify the first faulty action from partial trajectories without relying on future observations. 

\end{document}